\documentclass[11pt,a4paper]{article}
\usepackage[T1]{fontenc}
\usepackage{algorithm}
\usepackage{algpseudocode}
\usepackage{subcaption}

\usepackage[a4paper,top=2cm,bottom=2cm,left=3cm,right=3cm,marginparwidth=1.75cm]{geometry}
 
\usepackage{amsmath}
\usepackage{amssymb}
\usepackage{graphicx}
\usepackage{hyperref}

\usepackage[affil-it]{authblk}

\makeatletter
\renewcommand{\section}{\@startsection {section}{1}{\z@}%
              {24pt}{12pt} {\large\scshape\bfseries}}

\renewcommand{\subsection}{\@startsection {subsection}{2}{\z@}%
             {12pt}{12pt}  {\itshape\bfseries}}

\usepackage{apacite}
\usepackage{natbib}

\title{\bfseries \normalsize An Open-Source Tool for Reproducible Freeway Network Extraction from OpenStreetMap}

\author[1]{Drew Miller*}
\author[2]{Cathy Wu}

\affil[1]{Master's Student, Department of Civil and Environmental Engineering, Massachusetts Institute of Technology, United States}
\affil[2]{Associate Professor, Massachusetts Institute of Technology, United States}

\date{\vspace{-5ex}}

\begin{document}
\maketitle

\section*{Short summary}\small
Freeway simulation is often difficult to deploy at scale not only because of model formulation, but because preparing road network inputs remains a manual, corridor-specific, and difficult-to-reproduce task. This paper presents an \href{https://github.com/drewmiller-mit/osm2macrosim}{open-source tool} that extracts freeway networks from OpenStreetMap (OSM) and converts them into a compact, station-referenced representation suitable for downstream freeway simulation. Unlike existing tools that primarily support arterial or general network conversion tasks, the proposed workflow is designed around the specific requirements of freeway traffic studies. The tool supports not only OSM data cleaning and conversion, but also the broader workflow required in practice: corridor-specific querying, visual inspection of extracted segments, extraction validation against OSM, and source-data validation against aerial imagery. A locally hosted frontend allows users to define corridor-specific queries, select endpoints visually, and inspect extracted segments. 

The extraction logic is designed to address several recurring challenges in freeway OSM data, including inconsistent route references, ambiguous path selection through interchanges, managed-lane interference, incomplete corridor capture from naive bounding-box queries, and inconsistent ramp classifications. The workflow was first tested on two prototype corridors, where the extract-first-then-validate approach proposed here required roughly one-third the analyst effort of manual ramp encoding from scratch. It was then deployed across 359.6 miles of freeway in Orange County, California, with total processing and validation averaging about 41 seconds per mile. This deployment also suggests that, in a well-mapped region, OSM is sufficiently accurate for many freeway traffic studies. Overall, the tool provides a more scalable and reproducible foundation for freeway network preparation.

\textbf{Keywords}: freeway simulation, OpenStreetMap, reproducibility, road network data, traffic data processing

\section{Introduction}
Traffic simulation workflows differ substantially by roadway context. Arterial studies often focus on intersections, signal timing, turning movements, and localized driver behavior, making them well suited to microscopic simulation. Freeway studies, by contrast, are typically concerned with corridor-scale congestion dynamics such as bottleneck formation and queue propagation. These systems can be modeled microscopically or macroscopically, but macroscopic approaches are often especially attractive because they capture traffic-wave behavior at lower computational cost.

Deploying freeway simulation \emph{at scale}, however, remains difficult in practice, largely because preparing road network data is still highly manual. For each corridor, the analyst must identify the mainline extent, discretize the roadway, encode merges and diverges, and specify lane counts in a model-compatible format. Although doing so manually is manageable for one case study, this process becomes tedious, time-consuming, and difficult to reproduce when extended to many corridors or an entire regional freeway system. As a result, the scale of many freeway simulation studies is constrained by implementation effort.

Existing map-based tools only partially address this problem. \cite{ortmann_dyntapy_2022}, \cite{lu_virtual_2023}, \cite{qu_what_2024}, \cite{seo_uxsim_2023}, \cite{rapelli_tust_2019}, and \cite{lopez_microscopic_2018} have developed workflows to support importing OpenStreetMap (OSM) (\cite{openstreetmap_contributors_openstreetmap_2026}) data, but these tools cater more toward arterial simulation and dynamic traffic assignment (DTA) analyses. They address common OSM data-quality limitations for intersection-oriented applications, such as lane-level connectivity ambiguities. Freeway workflows, however, pose a different set of requirements. Realistic freeway simulation depends on accurate representation of ramps, lane drops, lane additions, and corridor continuity. In addition, available tools often assume that the study corridor has already been manually isolated from OSM, provide limited support for validation against aerial imagery, and frequently lack a graphical interface for efficient map-based inspection and editing. 

This work presents an open-source tool for extracting freeway networks from OSM and transforming them into a representation suitable for freeway simulation. The workflow automates mainline isolation, lane and ramp processing, and other roadway attributes needed for downstream modeling, while also supporting corridor-specific querying and structured validation through a frontend UI and generated HTML and KML artifacts. By supporting the broader extract-then-validate workflow rather than only map conversion, the proposed approach enables a more scalable, reproducible, and efficient process for freeway network preparation. The tool is available at \href{https://github.com/drewmiller-mit/osm2macrosim}{this GitHub repository}.


\section{Methodology}

\begin{figure}[H]
  \centering
  \includegraphics[width=\linewidth]{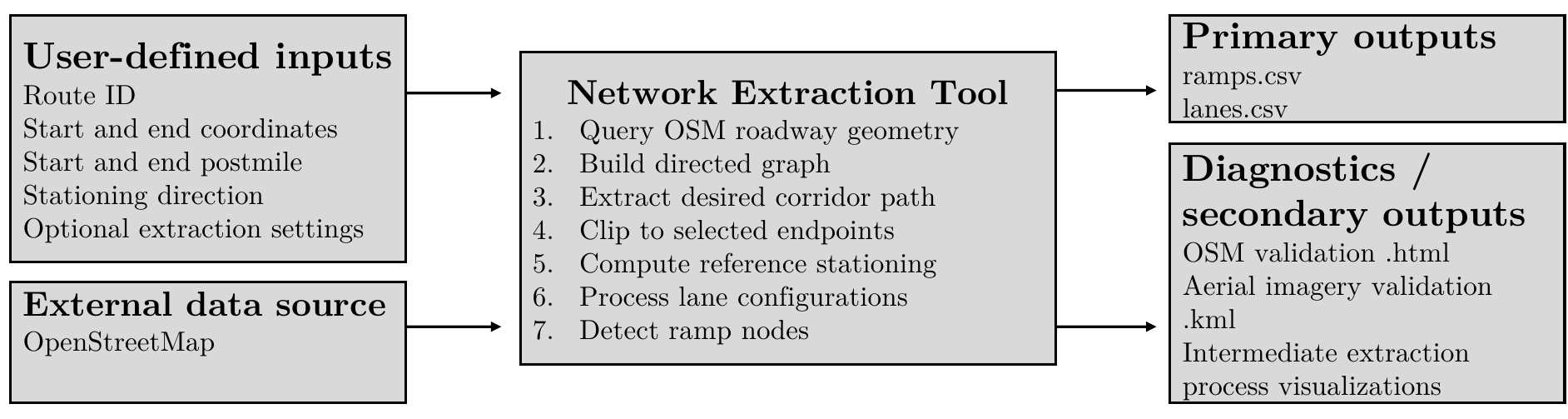}
  \caption{Inputs, outputs, and core processes in the network extraction tool.}
  \label{fig:network-black-box}
\end{figure}

As shown in Figure~\ref{fig:network-black-box}, the tool takes as input a route identifier, start and end coordinates, start and end postmiles, the stationing direction, and optional extraction settings, and combines these with roadway geometry obtained from OSM. Internally, it constructs a directed roadway graph, isolates the desired mainline path, clips the corridor to the selected limits, computes reference stationing, processes lane configuration changes, and detects ramp connection points. The primary outputs are \texttt{lanes.csv} and \texttt{ramps.csv}, which summarize lane configuration and ramp locations in a compact format. The tool also produces validation artifacts, including HTML and KML files, to support quality control before the extracted network is used in simulation or calibration workflows.

Several authors, including \cite{qu_what_2024} and \cite{meng_topology-preserving_2022}, have documented and addressed issues that arise when converting OSM arterial networks into simulation-ready road networks. Comparatively, data issues specific to freeway extraction are less documented. Table~\ref{tab:OSM_issues} summarizes several recurring issues that complicate freeway extraction from OSM, all of which this tool is designed to address.

\begin{table}[htbp]
\centering
\renewcommand{\arraystretch}{1.2}
\begin{tabular}{|p{3cm}|p{12cm}|}
\hline
\textbf{Recurring Issue} & \textbf{Description} \\
\hline
Inconsistency in mainline ref tags &
The same freeway corridor may be tagged with different route-reference variants, partial labels, concurrent route numbers, or missing \texttt{ref} values, making strict filtering unreliable. \\
\hline
Weaving managed/HOV/toll lanes &
Managed lanes may diverge from the general-purpose freeway, run in parallel, and later merge back, creating ambiguity about which path represents the intended mainline. \\
\hline
Bounding box overflow &
A query defined only by corridor endpoints can omit roadway sections in between them, especially on curved facilities. \\
\hline
Inconsistency in ramp classifications &
Ramps and connectors are not always tagged as \texttt{motorway\_link}, and some inflow/outflow points relevant to freeway modeling receive different classifications. \\
\hline
Ambiguous path selection through interchanges &
At complex interchanges, multiple directed paths may connect the same endpoints, and naive shortest-path selection may follow a connector or alternate branch rather than the intended mainline. \\
\hline
\end{tabular}
\caption[Recurring OSM-related issues for freeway studies]{Recurring OSM-related issues encountered when extracting freeway corridors for simulation workflows.}
\label{tab:OSM_issues}
\end{table}

\subsection{OSM Schema and Query Logic}

In OSM, a road is represented as an ordered \emph{way} composed of individual \emph{nodes}. Each way carries descriptive tags such as \texttt{highway}, \texttt{ref}, \texttt{lanes}, and \texttt{oneway}. For freeway extraction, the most important tags are \texttt{highway}, which identifies roadway class, and \texttt{ref}, which stores the signed route designation. The tool therefore operates primarily at the way level, while nodes preserve geometry and connectivity.

Given user-specified corridor endpoints, the tool first constructs a buffered bounding box and issues a query to the Overpass API. Rather than querying only the target freeway, the current implementation retrieves all ways within the bounding box whose \texttt{highway} tag is \texttt{motorway}, \texttt{motorway\_link}, or \texttt{primary}, along with the associated nodes needed to reconstruct geometry and topology. This broader query is necessary because, as stated above, freeway corridors are not always encoded as a clean, uninterrupted chain of \texttt{motorway} ways. Short connectors, auxiliary ramps, braided interchange links, and occasional nonstandard tagging can interrupt a stricter mainline-only query.

\begin{figure}[H]
    \centering
    \includegraphics[width=0.4\linewidth]{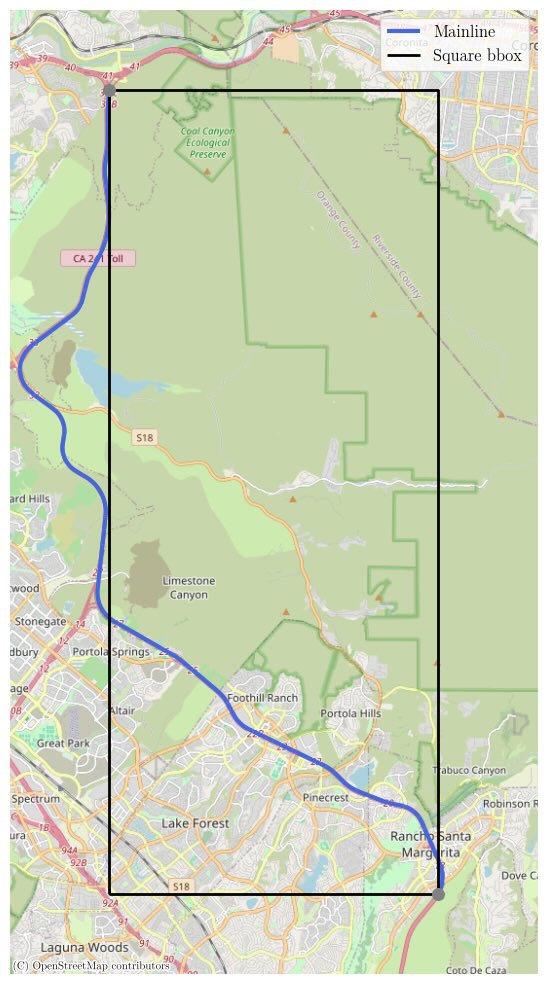}
    \caption[Bounding box queries]{A case where the simple bounding box query method misses important roadway sections.}
    \label{fig:bbox}
\end{figure}

The size of the bounding-box buffer is an important design choice. If the buffer is too small, the query may omit portions of the intended corridor when the roadway curves outside the rectangle implied by the selected endpoints. If the buffer is too large, the query becomes slower and returns more irrelevant geometry, increasing ambiguity during path extraction. Figure~\ref{fig:bbox} illustrates this issue for CA-241 northbound.

After the query is returned, the downloaded ways and nodes are converted into a directed graph. Edge direction is determined from the OSM \texttt{oneway} tag, and each edge retains attributes including way ID, route reference, roadway class, and geometric length. This graph forms the basis for corridor isolation.

\subsection{Mainline Isolation}

Once the roadway graph has been constructed from the queried OSM data, the next task is to isolate the corridor’s intended mainline path. This includes decisions such as travel direction and whether HOV or express lanes should be treated as part of the mainline. If the user does not explicitly provide OSM start and end node IDs, the tool infers them from the input coordinates by identifying nearby ways and selecting nodes with a directional bias that tends to extend the corridor slightly beyond the clicked location rather than truncating it prematurely. This helps avoid errors when user-selected endpoints fall in the middle of a way or near a complicated interchange.

To guide path extraction, the user-supplied freeway name is normalized into a regular-expression pattern matching common OSM \texttt{ref} tag variations. This is necessary because the same route may appear as \texttt{I 5}, \texttt{I-5}, \texttt{I5 S}, or as part of a longer combined reference string. The normalized regex provides a flexible way to identify candidate mainline edges even when tagging conventions vary.

Path extraction is then performed through a staged relaxation procedure. The tool first searches a tightly filtered subgraph containing only \texttt{motorway} edges whose \texttt{ref} tags match the normalized route regex. If no feasible directed path exists, the constraints are progressively relaxed: first by treating route-reference matches as a preference rather than a requirement, then by broadening roadway classes, and finally by searching the full queried graph while still favoring route-reference matches. This staged logic allows the tool to recover valid freeway corridors even when short segments are missing \texttt{ref} tags, temporarily mapped with a different roadway class, or encoded with inconsistent metadata. Algorithm~\ref{alg:mainline_isolation} provides psuedocode for this mainline isolation process. 

\begin{algorithm}[htbp]
\caption{Mainline Isolation}
\label{alg:mainline_isolation}
\begin{algorithmic}[1]
\Require freeway name \texttt{interstate\_name}; start coordinate $(\lambda_s,\phi_s)$; end coordinate $(\lambda_e,\phi_e)$; directed graph \texttt{G}; optional \texttt{start\_osm\_node}; optional \texttt{end\_osm\_node}; optional \texttt{path\_mode}; optional \texttt{ref\_list}
\Ensure ordered mainline way IDs \texttt{way\_ids}; ordered path nodes \texttt{path\_nodes}

\State \texttt{interstate\_regex} $\gets$ \Call{NormalizeInterstateName}{\texttt{interstate\_name}}

\If{\texttt{start\_osm\_node} is not provided or \texttt{end\_osm\_node} is not provided}
    \State \texttt{start\_osm\_node}, \texttt{end\_osm\_node} $\gets$ \Call{InferPathNodes}{\texttt{result}, $(\lambda_s,\phi_s)$, $(\lambda_e,\phi_e)$}
\EndIf

\State \texttt{G\_1} $\gets$ \Call{FilteredSubgraph}{\texttt{G}, highway class $=$ \texttt{motorway}, ref match $=$ \texttt{interstate\_regex}, require match $=$ true}
\State \texttt{G\_2} $\gets$ \Call{FilteredSubgraph}{\texttt{G}, highway class $=$ \texttt{motorway}}
\State \texttt{G\_3} $\gets$ \Call{FilteredSubgraph}{\texttt{G}, ref match $=$ \texttt{interstate\_regex}, require match $=$ true}

\State \texttt{stages} $\gets$ [
\Statex \hspace{\algorithmicindent} $(\texttt{G\_1}, \texttt{normal}, \varnothing, \texttt{regex})$,
\Statex \hspace{\algorithmicindent} $(\texttt{G\_2}, \texttt{prefer}, [\texttt{interstate\_regex}], \texttt{regex})$,
\Statex \hspace{\algorithmicindent} $(\texttt{G\_3}, \texttt{normal}, \varnothing, \texttt{exact})$,
\Statex \hspace{\algorithmicindent} $(\texttt{G}, \texttt{prefer}, [\texttt{interstate\_regex}], \texttt{regex})$
\Statex \hspace{\algorithmicindent} ]

\For{each stage $(\texttt{G\_stage}, \texttt{mode}, \texttt{refs}, \texttt{match\_mode})$ in \texttt{stages}}
    \If{\Call{DirectedPathExists}{\texttt{G\_stage}, \texttt{start\_osm\_node}, \texttt{end\_osm\_node}}}
        \State \texttt{way\_ids}, \texttt{path\_nodes} $\gets$ \Call{SeparatePath}{\texttt{G\_stage}, \texttt{start\_osm\_node}, \texttt{end\_osm\_node}, \texttt{refs}, \texttt{mode}, \texttt{match\_mode}}
        \State \textbf{break}
    \EndIf
\EndFor

\If{\texttt{way\_ids} is empty}
    \State \textbf{raise error} ``No feasible directed mainline path found''
\EndIf

\State \Return \texttt{way\_ids}, \texttt{path\_nodes}
\end{algorithmic}
\end{algorithm}

The tool also supports optional user-defined \texttt{prefer} and \texttt{avoid} modes, in which selected route references are explicitly favored or penalized during path selection. This is especially useful when unwanted express or managed lanes satisfy the same regex filter as the general-purpose freeway. Once selected, the mainline is clipped precisely to the user-specified corridor limits and passed to downstream processing.

\subsection{Reference Stationing, Lane Processing, and Ramp Detection}

After the correct mainline has been identified, the remaining steps convert that geometry into a station-referenced representation suitable for simulation. The user provides starting and ending postmile values, and cumulative distance along the clipped mainline is linearly scaled to match that postmile range. This prevents small geometric discrepancies in OSM from accumulating into downstream stationing drift.

Lane processing then converts raw OSM \texttt{lane} tags into a compact sequence of station-referenced lane blocks. Segments with valid lane counts are projected onto the reference line, ordered longitudinally, and merged where adjacent segments share the same lane count. The result is a simplified profile of lane additions and drops along the corridor.

Ramp processing identifies mainline nodes that function as inflow or outflow points by examining connected OSM ways that are not part of the extracted mainline. Importantly, this logic does not rely solely on the \texttt{motorway\_link} tag, since ramps and other relevant connectors are not always classified consistently in OSM. Nodes are classified according to whether connected ways begin or end at the mainline node, consistent with stored way direction. Retained ramp nodes are then projected onto the station-referenced corridor.

\subsection{GMNS Interoperability and Validation}

Near the end of the extraction pipeline, the processed corridor can also be exported to General Modeling Network Specification (GMNS) format, producing \texttt{node.csv}, \texttt{link.csv}, and \texttt{config.csv}. This representation differs from \texttt{lanes.csv} and \texttt{ramps.csv}: the latter are compact station-based abstractions intended for freeway simulation, while the GMNS files encode a general-purpose directed network for interoperability with external transportation tools.

The tool also includes a frontend for interactive use. It allows users to select corridor endpoints visually, configure runs, inspect intermediate artifacts, and review outputs without interacting directly with code.

\begin{figure}[H]
    \centering
    \begin{subfigure}[tbp]{0.28\linewidth}
        \includegraphics[width=\linewidth]{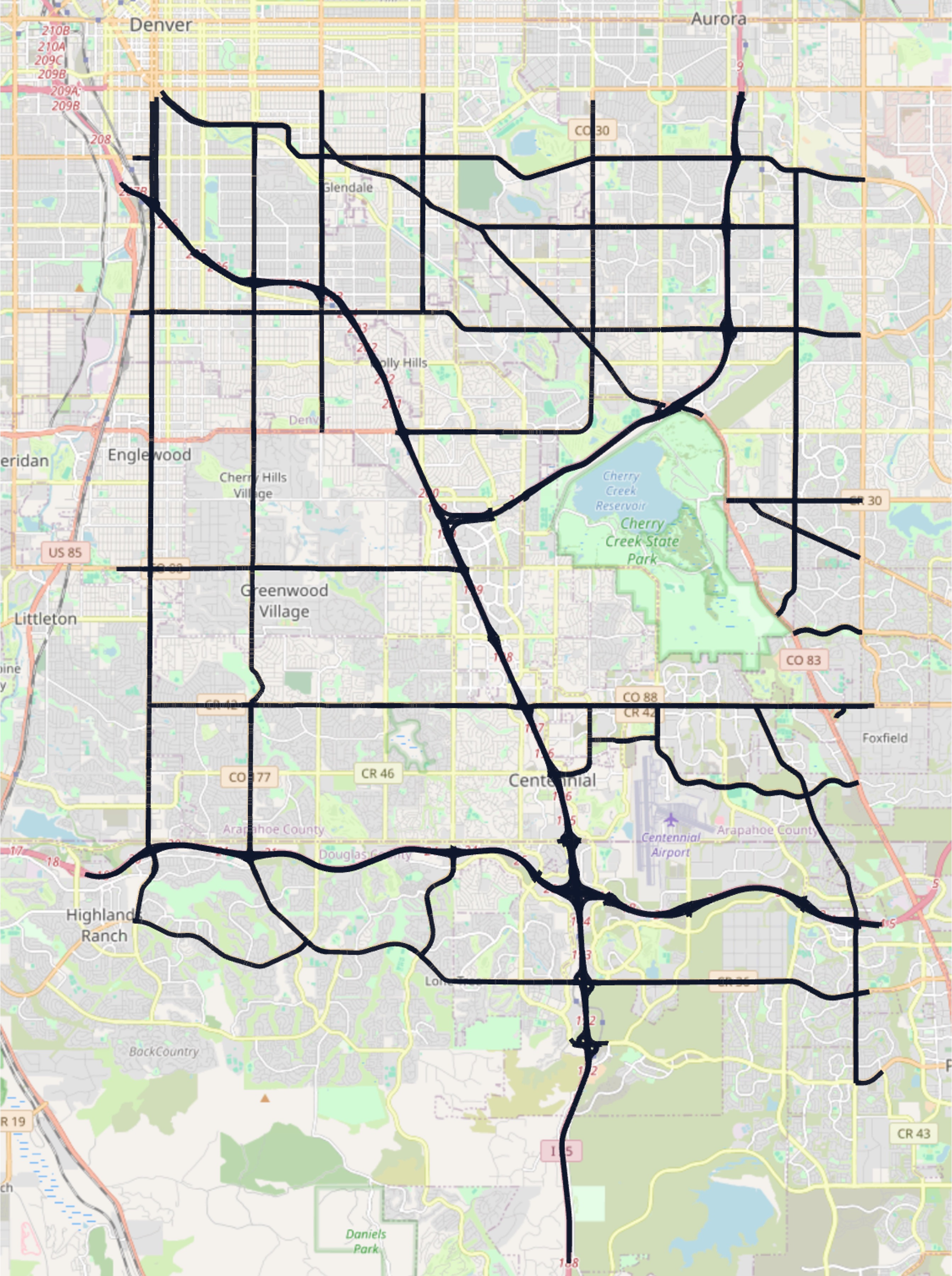}
        \label{}
    \end{subfigure}
    \hfill
    \begin{subfigure}[tbp]{0.18\linewidth}                          \includegraphics[width=\linewidth]{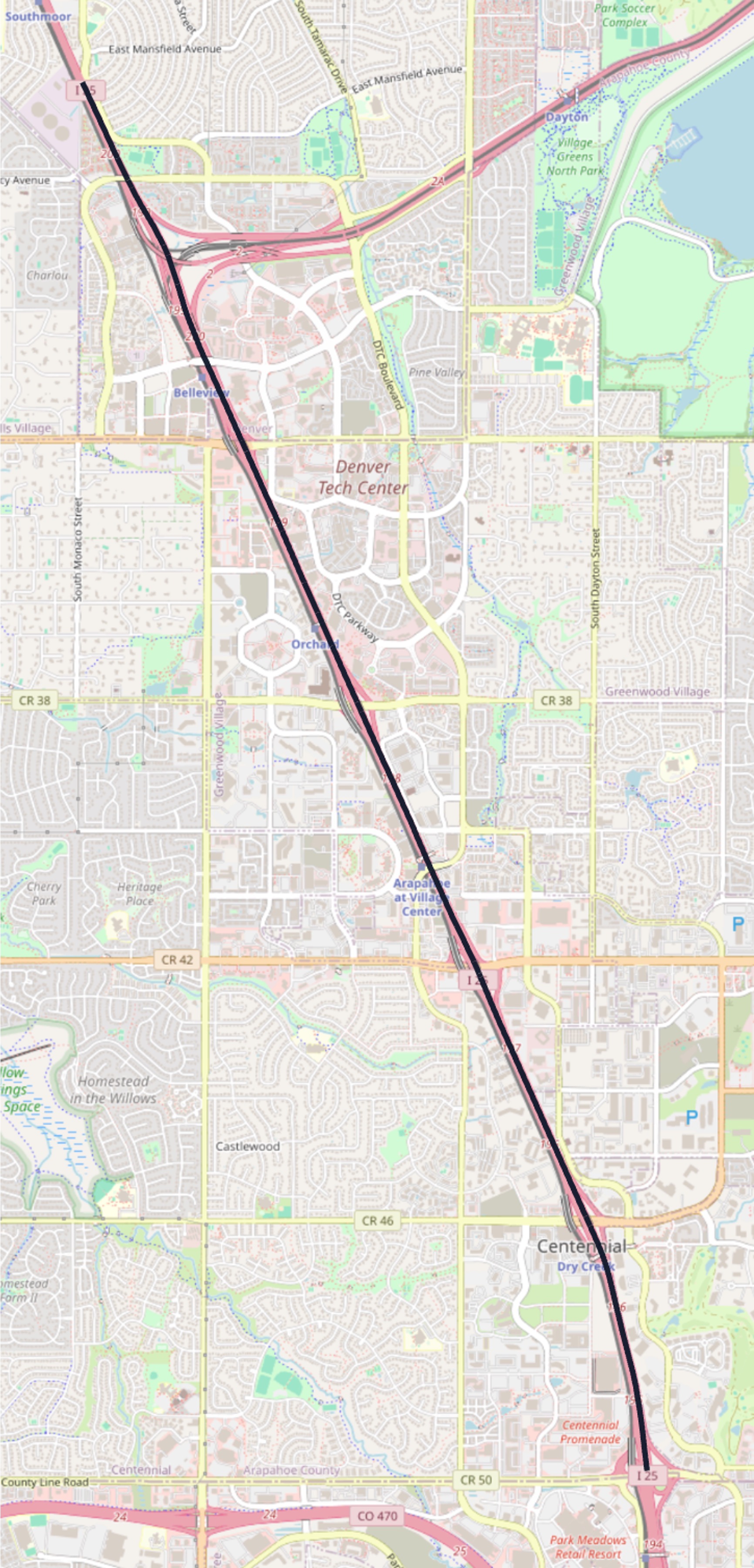}
            \label{}
    \end{subfigure}
    \hfill
    \begin{subfigure}[tbp]{0.21\linewidth}        \includegraphics[width=\linewidth]{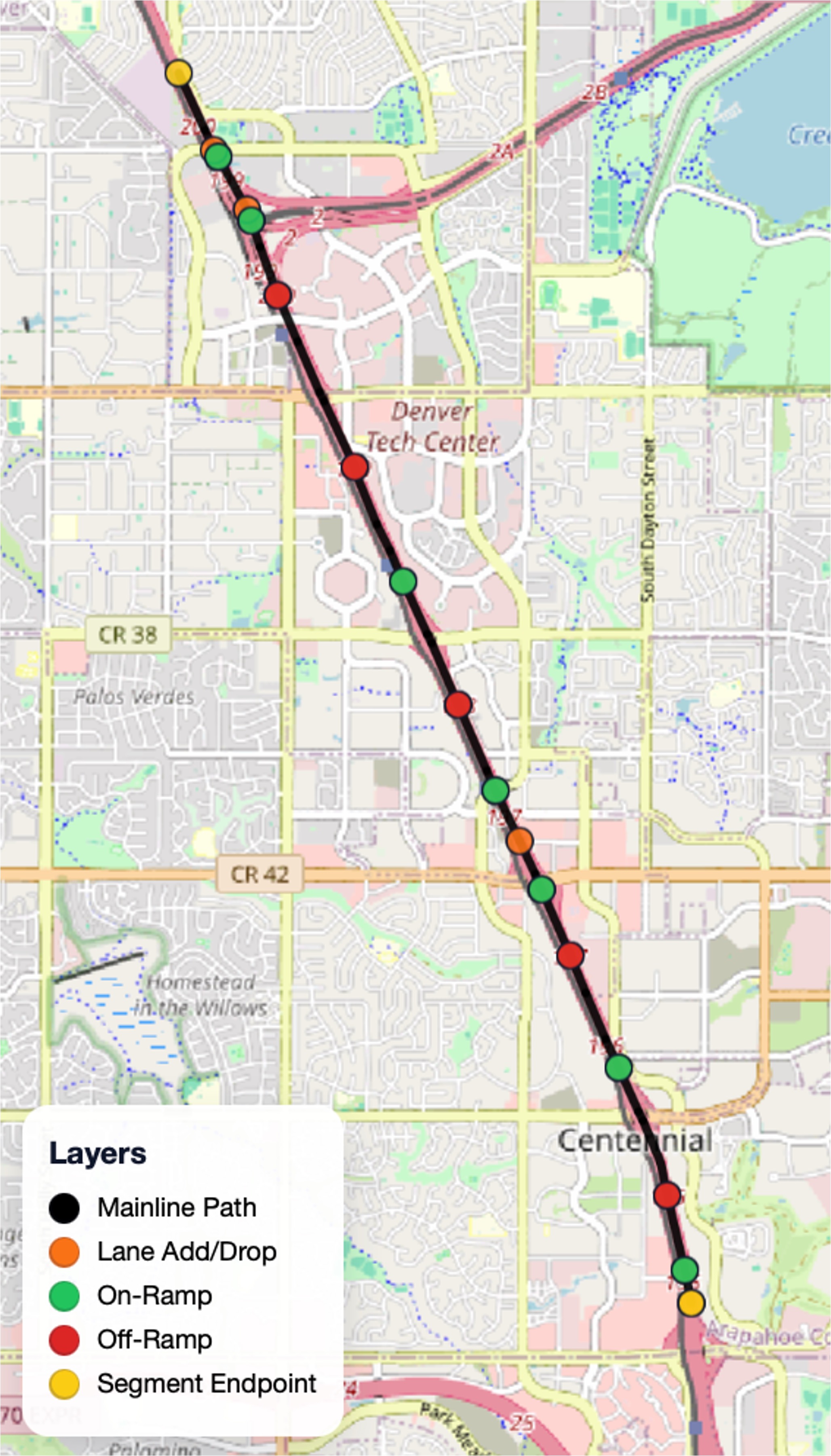}
            \label{}
    \end{subfigure}
    \hfill
    \begin{subfigure}[tbp]{0.28\linewidth}        \includegraphics[width=\linewidth]{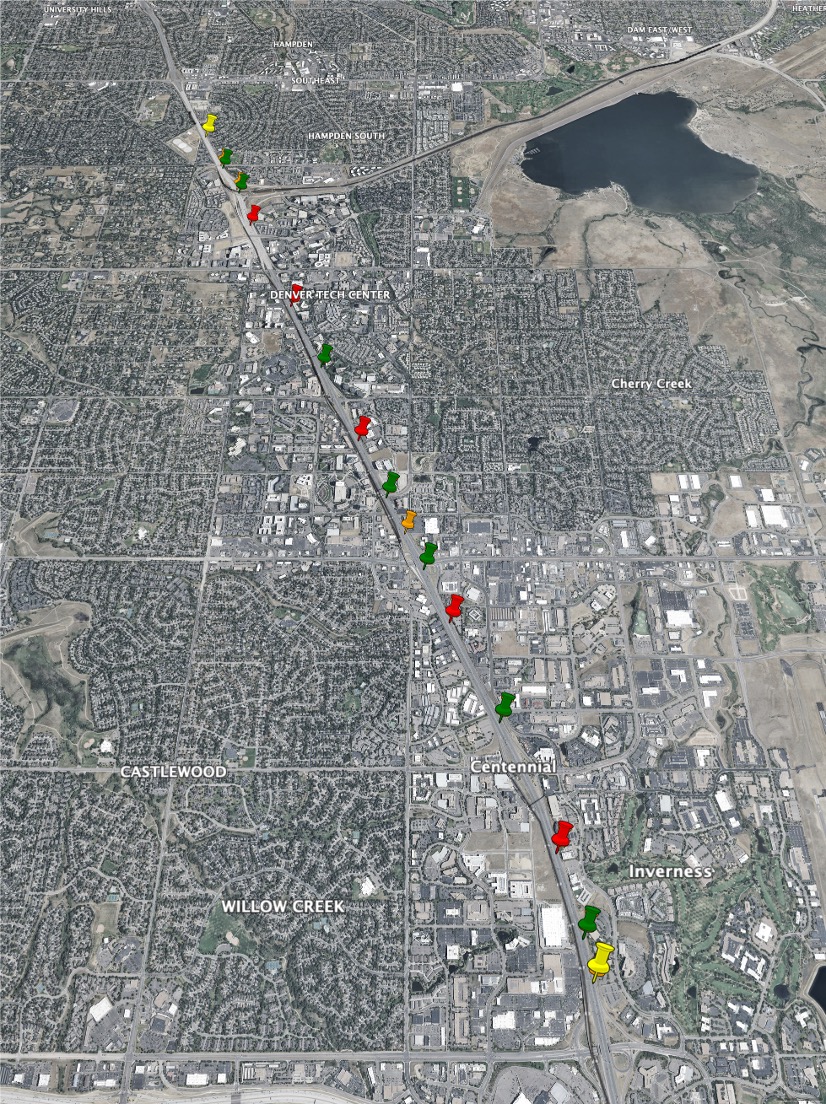}
    \label{}
    \end{subfigure}
    \caption{From left to right, visualizations of the tool's initial OSM query, the mainline isolation, the extraction validation, and the source-data validation.}
    \label{fig:workflow}
\end{figure}

\begin{figure}[H]
    \centering
    \begin{subfigure}[tbp]{0.7\linewidth}
        \includegraphics[width=\linewidth]{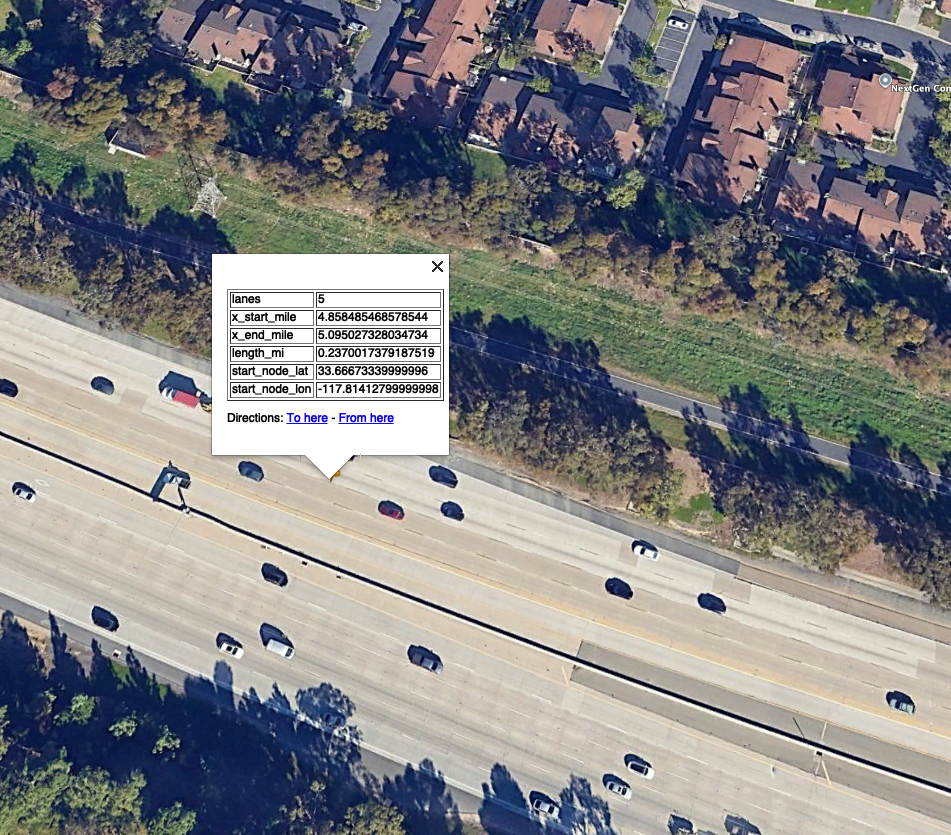}
    \end{subfigure}
    \begin{subfigure}[tbp]{0.7\linewidth}
        \includegraphics[width=\linewidth]{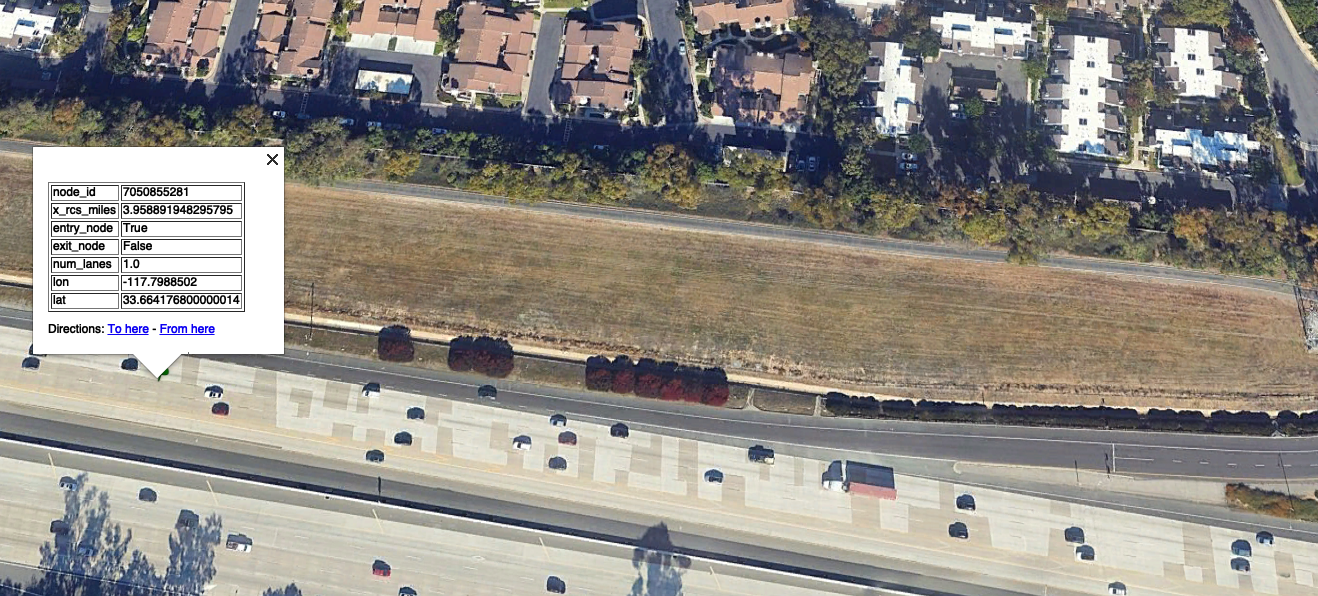}
    \end{subfigure}
    \caption{Examples of source-data validation to verify the presence of a lane addition (top) and an inflow point (bottom).}
    \label{fig:ramp_lane}
\end{figure}

Finally, the tool generates two validation artifacts. The first is an HTML map showing the extracted mainline, lane changes, and ramp nodes over the queried OSM network. This supports \emph{extraction validation}, helping determine whether the tool interpreted OSM as intended. The second is a KML artifact that can be overlaid on aerial imagery in Google Earth. This supports \emph{source-data validation}, allowing the user to assess whether OSM itself is an adequate representation of the real-world roadway. Together, these artifacts provide a practical two-stage workflow: first verify the extraction logic, then verify the underlying source data. Figure~\ref{fig:workflow} shows artifacts of the full extract-then-validate workflow supported by this tool, and Figure~\ref{fig:ramp_lane} displays zoomed-in screen captures of source-data validation examples using aerial imagery.

\subsection{Evaluation Design}
After defining the extraction workflow and its outputs, the next step is to assess how well the tool performs in practice and how reliable the underlying OSM data is for freeway applications. Several tests were conducted to assess three aspects of the tool itself and OSM data quality: first, the analyst time savings achieved relative to a more conventional manual encoding process; second, the tool’s ability to extract freeway corridors at scale; and third, the suitability of OSM road network data for freeway simulation. To support these evaluations, the tool was first benchmarked against manual extraction on two test corridors, I-24 in Nashville, Tennessee, and I-35W in Dallas, Texas. It was then deployed across 360 miles of freeway in Orange County, California.
\section{Results and discussion}
Before deploying the tool across Orange County, California, it was prototyped on two freeway corridors outside the final study area. These early case studies served as proof-of-concept deployments and were used to assess both whether the extraction logic produced usable freeway networks and whether the overall extract-first-then-validate workflow was faster than constructing equivalent networks manually from aerial imagery.

\begin{figure}
    \centering
    \includegraphics[width=0.3\linewidth]{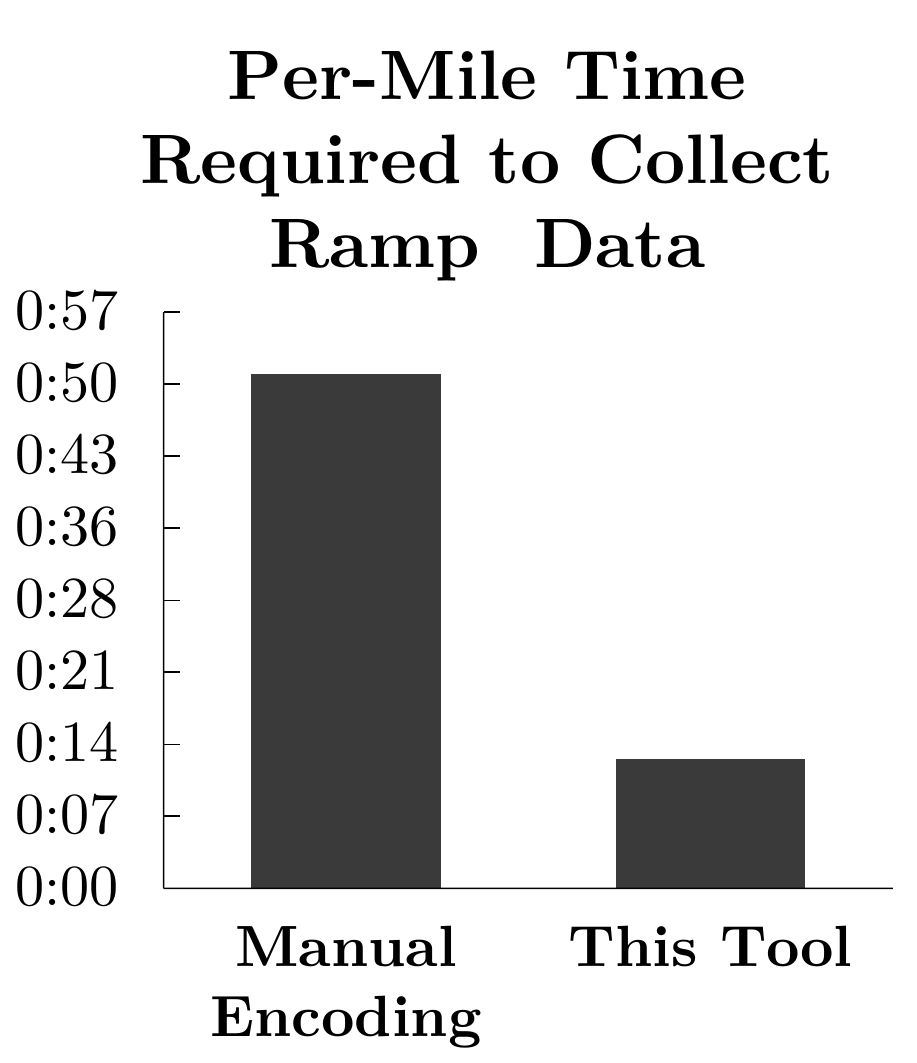}
    \caption{The tool presented in this paper reduces analyst time by approximately three times.}
    \label{fig:collection_time}
\end{figure}

For these early tests, the time comparison focused specifically on ramp-feature collection rather than the full lane-and-ramp network. The comparison measured the time required to validate automatically extracted ramp locations against imagery versus the time required to manually encode those same ramp features from scratch in Google Earth. As shown in Figure~\ref{fig:collection_time}, the extract-first workflow substantially reduced analyst effort. On a per-mile basis, manual ramp encoding required roughly three times more time than extraction followed by validation enabled by this tool. This finding motivated the subsequent district-scale deployment in Orange County, where the goal shifted from proof of concept to evaluating performance across a broader inventory of freeway corridors.

\subsection{Orange County, California Network Extraction}

\begin{figure}[tbp]
    \centering
    \begin{subfigure}[t]{0.46\textwidth}
        \centering
        \includegraphics[width=\linewidth]{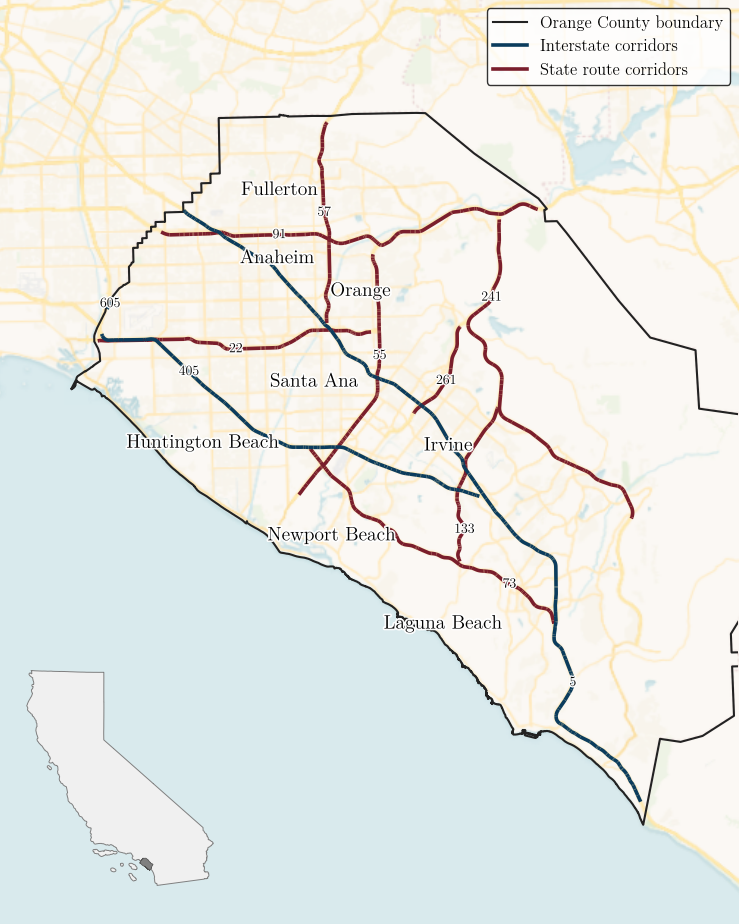}
        \caption[District 12 corridors]{All extracted freeway corridors are within the CalTrans Performance Measurement System (PeMS) District 12 study boundary.}
        \label{fig:d12}
    \end{subfigure}
    \hfill
    \begin{subfigure}[t]{0.52\textwidth}
        \centering
        \includegraphics[width=\linewidth]{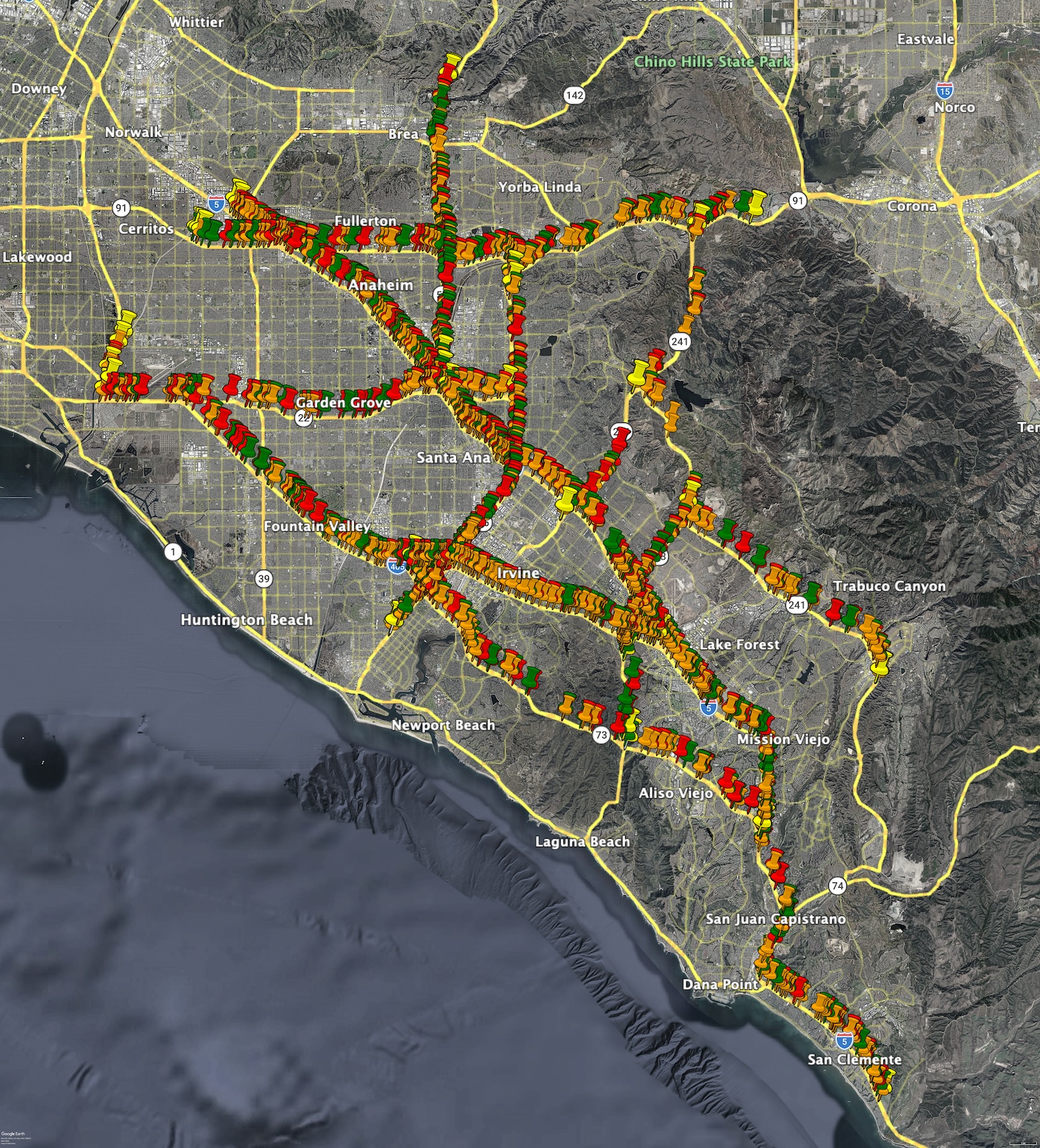}
        \caption[Orange County aerial validation]{Extracted geometric attributes of interest in Orange County overlaid on aerial imagery for validation. Here, green markers correspond to inflow points, red to outflow points, orange to lane add/drop locations, and yellow to corridor start/end points.}
        \label{fig:d12_aerial}
    \end{subfigure}
\end{figure}

Figure~\ref{fig:d12} summarizes the resulting network inventory from this county-wide deployment. To initialize each run, the first and last mainline detector stations on each distinct freeway were identified from the \cite{caltrans_highway_2026} highway Performance Measuring System (PeMS) District~12 station inventory, and their coordinates were used as the start and end points for the extraction pipeline. Because these detector stations also include postmile information, they provided a natural way to define both the geometric corridor endpoints and the stationing range used for each extraction.

After each run completed, the first validation step was a quick review of the interactive OSM HTML artifact to confirm that the extracted path followed the intended mainline. In two corridors, \texttt{I-405} and \texttt{I-5}, this step revealed that manual intervention was needed: route references associated with express or HOV facilities were specified using \texttt{avoid} mode so that the extracted path would remain on the general-purpose mainline. Once the path was verified against OSM, a second validation step was carried out using the KML outputs overlaid on aerial imagery, as shown in Figure~\ref{fig:d12_aerial}. This stage required more time because it involved checking whether OSM lane-change and ramp information matched the real-world roadway.

In total, the District~12 validation effort, excluding the time required to submit jobs and download artifacts, required approximately 245 minutes and produced 359.6 miles of extracted freeway roadway. This corresponds to an average processing and validation effort of about 41 seconds per mile. These results suggest that corridor-scale freeway extraction can be deployed across a moderately large urban district within a reasonable amount of analyst time, even when manual validation is retained as part of the workflow.

\subsection{Observed OSM Error Patterns}

Lane-related issues were the most common OSM data problem encountered during validation, with 20 total lane errors across all extractions. Importantly, this was not due to missing lane-count data: lane attributes were present on the extracted OSM ways throughout the network. Instead, the issue was that the reported lane counts were occasionally incorrect relative to aerial imagery. By contrast, ramp-related errors were uncommon. Only three instances required adding or deleting a ramp solely because it was represented incorrectly in OSM. This suggests that the underlying freeway topology in OSM is generally reliable, even when some operational attributes are imperfect.

A more frequent issue was that the exact merge or diverge point represented in OSM did not align perfectly with the location suggested by aerial imagery. For freeway macrosimulation, however, some positional tolerance is acceptable because an inflow or outflow only needs to be assigned to the correct spatial unit. In other words, the significance of this type of positional error depends on the model discretization: the larger the spatial step, the more tolerance the workflow has for slight offsets in ramp location.

Taken together, these validation results indicate that manual review remains important, particularly because lane attributes can be inaccurate even when they are available everywhere in the network. At the same time, the Orange County deployment suggests that OSM data quality is already strong enough for many freeway use cases. This is an important distinction from more geometry-sensitive simulation workflows, where lane-level detail, exact merge points, and other fine-grained roadway controls often require greater precision and more extensive manual correction.



\section{Conclusions}
By combining automated extraction with structured validation artifacts, the proposed tool helps shift the freeway network data processing workflow from ad hoc manual encoding toward a more transparent and reusable pipeline.

At the same time, several limitations remain. Most importantly, the workflow still depends on manual validation against aerial imagery and OSM itself. This review step was manageable in the present county-level study, but it remains a bottleneck for larger deployments. A promising direction for future work is to further automate validation using vision-language agents that can compare extracted network features against imagery and flag likely discrepancies for analyst review. Such a human-in-the-loop approach could preserve quality control while further reducing labor.

A second limitation is geographic generalizability. The Orange County case study suggests that OSM is already of sufficiently high quality for many freeway simulation tasks in a dense, well-mapped urban region. However, this finding may not extend to less populous or less actively mapped areas, where OSM contribution levels may be lower and roadway attributes may be less complete or less reliable. Additional evaluation across rural areas and other regions is therefore needed before making broader claims about transferability.


\bibliography{references}

\end{document}